\documentclass[11pt,tightenlines]{revtex4-1}
\usepackage{graphics,graphicx,epsfig}
\usepackage{color}
\begin{document}
\title{Transition from compact to porous films in deposition with
temperature activated diffusion}
\author{Dung di Caprio${}^{1,}$\footnote{Email address: dung.di-caprio@chimie-paristech.fr}\\
F. D. A. Aar\~ao Reis${}^{2,}$\footnote{Email address: reis@if.uff.br}
}
\affiliation{
{(1) PSL Research University, Chimie ParisTech - CNRS, Institut de Recherche
de Chimie Paris, 75005, Paris, France}\\
(2) Instituto de F\'\i sica, Universidade Federal Fluminense,\\
Avenida Litor\^anea s/n, 24210-340 Niter\'oi RJ, Brazil}
\date{\today}

\begin{abstract}

We study a thin film growth model with temperature activated diffusion
of adsorbed particles, allowing for the formation of overhangs and pores, but
without detachment of adatoms or clusters from the deposit.
Simulations in one-dimensional substrates are performed for several values of
the diffusion-to-deposition ratio $R$ of adatoms with a single bond and of
the detachment probability $\epsilon$ per additional nearest neighbor (NN),
respectively with activation energies are $E_s$ and $E_b$.
If $R$ and $\epsilon$ independently vary, regimes of low and high porosity
are separated at $0.075\leq\epsilon_c\leq 0.09$,
with vanishingly small porosity below that point and finite porosity for larger
$\epsilon$.
Alternatively, for fixed values of $E_s$ and $E_b$ and varying temperature,
the porosity has a minimum at $T_c$, and a nontrivial regime in which it increases
with temperature is observed above that point.
This is related to the large mobility of adatoms, resembling features
of equilibrium surface roughening.
In this high-temperature region, the deposit has the structure of a critical percolation
cluster due to the non-desorption. 
The pores are regions enclosed by blobs of the corresponding percolating backbone,
thus the distribution of pore size $s$ is expected to scale as $s^{-\tilde\tau}$ with
$\tilde\tau\approx 1.45$, in reasonable agreement with numerical estimates.
Roughening of the outer interface of the deposits suggests Villain-Lai-Das Sarma
scaling below the transition. Above the transition, the roughness exponent
$\alpha\approx 0.35$ is consistent with the percolation backbone structure via
the relation $\alpha = 2-d_B$, where $d_B$ is the backbone fractal dimension.

\end{abstract}

\maketitle

\section{Introduction}
\label{intro}

Porous materials attract much interest due to their broad range of commercial and
scientific applications \cite{barton,hilfer,xuan,innocenzi}. Tailoring of materials
with the desired structure, physical, and chemical properties has been mainly an
empirical work. However, modeling the effects of growth conditions on
those properties may be a useful tool to improve them.
For instance, for applications of porous films of oxides or silicon, the simultaneous
regulation of porosity and surface roughness is suggested in Refs.
\protect\cite{huang2012,huang2013}.

In the study of thin porous film, the simplest models probably are
ballistic deposition (BD) \cite{vold,barabasi} and its extensions
\cite{pellegrini,kikkinides,bbd,perez}.
In these models, the surface relaxation processes take place during a short time
interval after the aggregation of incident particles (atoms, molecules, or clusters).
Changes of growth parameters may lead to drastic changes in the porosity and pore
connectivity of the deposits, typically with the porosity decreasing as surface
relaxation processes are enhanced \cite{pellegrini,yu,bbdflavio,khanin,banerjee}. 

Compact structures are usually observed in film growth dominated
by surface diffusion, particularly if diffusion lengths of adsorbed species are large
\cite{barabasi,etb}.
Some models with activated diffusion of adsorbed species may produce porous deposits
at low temperatures, but the porosity decreases as the substrate temperature increases
\cite{hu2009,zhang2010}.
In experiments and models, the formation of larger crystalline grains and smoother
film surfaces are also observed as the substrate temperature increases \cite{ohring}.

The opposite trend is observed in some systems, such as electrostatic spray
deposition of oxides \cite{chen,marinha}: transitions from compact to ramified
structures occur as the temperature increases.
Other physico-chemical parameters are also related to these features, such as the liquid
content of the spray droplets.
For instance, solution evaporation and clustering in those droplets before adsorption
is suggested to explain the formation of ramified structures \cite{chen}.
However, the results also suggest to investigate whether surface diffusion
of adsorbed atoms or molecules may lead to increase of film porosity or roughness.

In this paper, we introduce a model of thin film deposition with no desorption and
with temperature activated adatom diffusion that shows this trend at high temperatures.
It is an extension of the Clarke-Vvedensky (CV) model for thin film growth \cite{etb,cv}
without the usual solid-on-solid condition and without desorption.
For fixed value of binding energies, there is
a (low temperature) regime with porosity decreasing with temperature and, above
a transition point, a regime with porosity increasing with temperature.
In the latter regime, the deposit has a critical percolation backbone structure,
which is reflected in the roughness scaling, and
the pore size distribution has a power-law decay related to
percolation exponents. This suggests an alternative interpretation for the increase
of porosity in high temperature deposition processes and the possibility of producing
critical pore size distributions in a broad range of growth parameters.

The rest of this paper is organized as follows. In Sec. \ref{model}, we present the
deposition model and the simulation procedure.
In Sec. \ref{porous}, we analyze the internal structure of the deposits,
focusing on the changes of porosity as the model parameters change.
In Sec. \ref{poredistribution}, we present the pore size distributions obtained
in simulations and explain their high-temperature decay using percolation concepts.
 In Sec. \ref{roughness}, we discuss the surface roughness scaling in the deposits.
In Sec. \ref{conclusion}, we summarize our results and present our conclusions.

\section{Model and simulation procedure}
\label{model}

Deposition occurs in a one-dimensional substrate of linear size $L$,  with the lattice
constant taken as the length unit. Periodic boundary conditions are considered
in the lateral direction ($x$).
The surface is flat at $t=0$, with all columns with height $z=0$.

There is an external flux of $F$ atoms per site per unit time.
The incident atom is adsorbed upon landing above a previously deposited atom or a
substrate site.
After adsorption, it executes activated diffusion steps with rules described below.
There is no relaxation to lower heights of neighboring columns immediately after
adsorption, in contrast to the models of Refs. \protect\cite{hu2009,zhang2010,sanchez}.

\begin{figure}[!h]
\includegraphics[width=5cm]{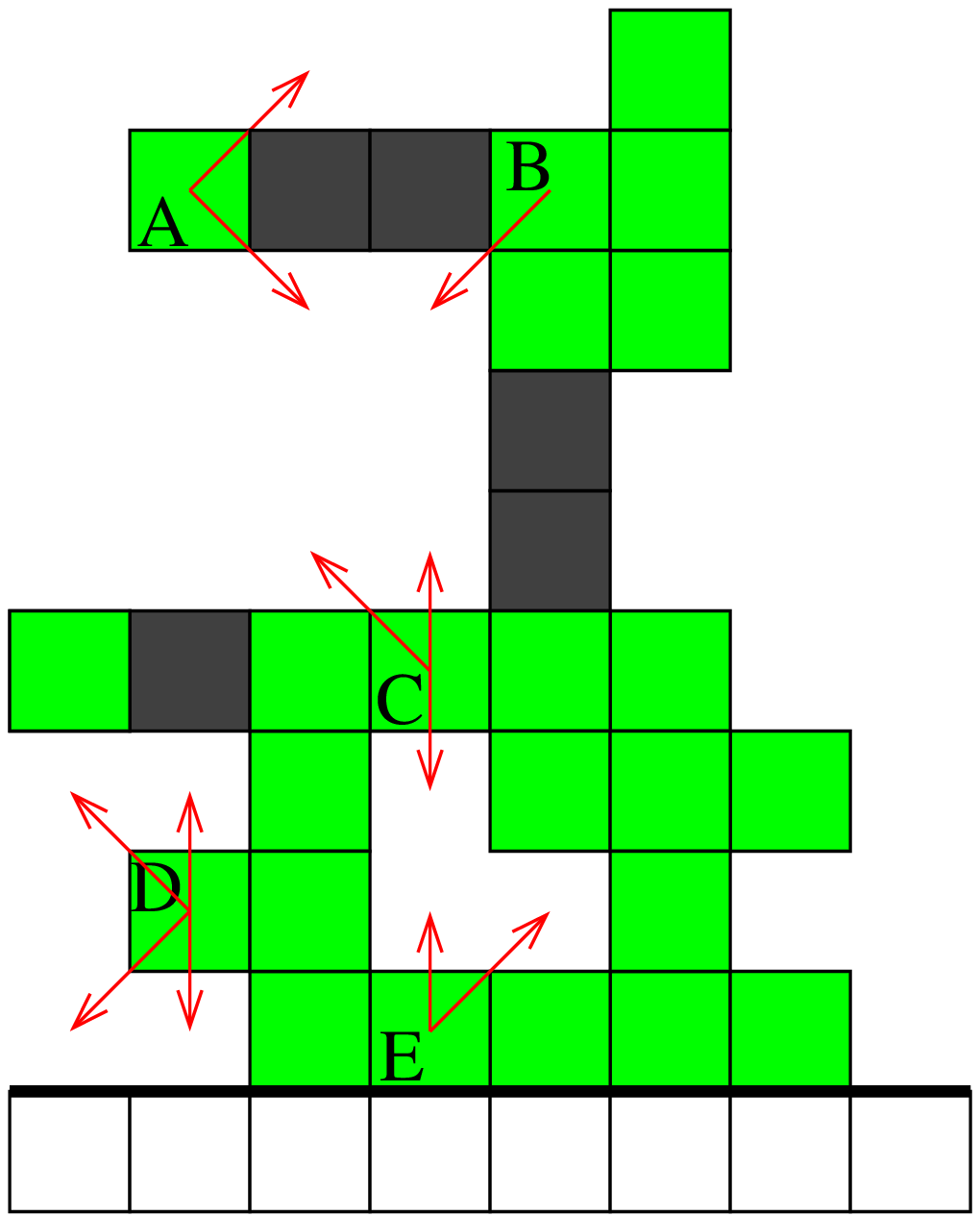}
\caption{(Color online) Schematic representation of the growing layer: substrate
sites are empty squares, mobile adatoms are colored (green) squares, and non-mobile
adatoms are black squares.
The allowed steps of five adatoms, labeled A to E, are indicated by red arrows.
}
\label{steps}
\end{figure}

Fig. \ref{steps} highlights the adatoms that are allowed to move in a given configuration
of the deposit and shows the possible steps of five of them.

The hopping rate (diffusion coefficient) ${\cal D}$ of an adatom at a given position
depends only on the configuration of the four nearest neighbor (NN) sites of this position:
$\left( x+1,y\right)$, $\left( x-1,y\right)$, $\left( x,y-1\right)$, and $\left( x,y+1\right)$.
That rate does not depend on the configuration of the target position
(i. e. the position after the step). Its general form is
\begin{equation}
{\cal D}=D\epsilon^{n-1} ,
\label{difcoef}
\end{equation}
where $n$ is the number of occupied NN sites and $\epsilon <1$ because NN bonds
reduce the adatom mobility.
Due to the choice of length unit, ${\cal D}$ is the number of random steps of the adatom
per unit time. Conversely, the average time for a step of this adatom is
\begin{equation}
\tau = 1/{\cal D} .
\label{tauD}
\end{equation}

In Fig. \ref{steps}, adatoms A and D have $n=1$ (${\cal D}=D$), adatom C has $n=2$
(${\cal D}=D\epsilon$), and adatoms B and E have $n=3$ (${\cal D}=D\epsilon^2$).
Note that a site of the substrate is included in the neighborhood of E.

The target position of the adatom step is randomly chosen among the NN and the
next nearest neighbor (NNN) sites; the latter option includes positions
$\left( x+1,y+1\right)$, $\left( x-1,y-1\right)$, $\left( x+1,y-1\right)$, and
$\left( x-1,y+1\right)$.
The target site may be in a height above or below the current level of the adatom.
The step is executed if the target site is empty and if a constraint is satisfied:
all adatoms remain connected to the substrate by a set of NN adatoms after the step.
This constraint prevents adatom desorption, which is a reasonable approximation for
many vapor growth processes.

In Fig. \ref{steps}, the adatoms in black cannot move because any step to an empty
NN or NNN site would disconnect them or disconnect other adatoms from the substrate.
The other adatoms are mobile, but some steps to NN or NNN sites are not allowed due
to the non-desorption condition.
Fig. \ref{steps} also shows the allowed steps of five labeled adatoms.
The case of adatom B is particularly interesting: if it is removed, a part of
the deposit is disconnected, but if it moves down-left, connectivity is restored.
For this reason, this is the only acceptable step of adatom B.

The diffusion coefficient of an adatom with a single NN is related to the temperature as
\begin{equation}
D = h\exp{\left( -E_s/k_BT\right)} ,
\label{Es}
\end{equation}
where $E_s$ is an energy  barrier and $h$ is a characteristic frequency
chosen as $h={10}^{12} s^{-1}$.
The reduction factor for additional NN is
\begin{equation}
\epsilon=\exp{\left( -E_b/k_BT\right)} ,
\label{Eb}
\end{equation}
where $E_b$ is a bond energy.
These rules are the same of the solid-on-solid CV
model in square or simple cubic lattice \cite{cv,ratsch1994}.
However, in contrast to the original CV model, here the adatom is also allowed to step
to empty NNN sites and the target position may be at any height, which leads to
overhang formation.

In the CV model, $E_s$ is interpreted as an interaction energy between the adatom
and the atomic layers below it, while $E_b$ is the interaction energy with
in-plane (lateral) neighbors.
Our model allows the formation of surface overhangs and pores,
thus many adatoms do not have atomic layers below them; this is the case of adatoms A, C,
and D in Fig. \ref{steps}.
For this reason, here $E_s$ is interpreted as a result of interactions of the adatom with
a large surrounding region, possibly including distant neighbors.
On the other hand, $E_b$ is still interpreted as a binding energy per additional NN.
This justifies the use of different values for $E_s$ and $E_b$, similarly to the
CV model \cite{cv,ratsch1994}.

The relative effects of diffusion and deposition are represented by the ratio
\begin{equation}
R\equiv \frac{D}{F} =\frac{h}{F} \exp{\left( -E_s/k_BT\right)} .
\label{defR}
\end{equation}
Thus, $R$ and $\epsilon$ are taken as the model parameters for our simulations.

Simulations of the model with large values of $R$ (of order ${10}^9$ or more) are usual in
submonolayer growth studies \cite{shim,tiagoreversible},  but demand long simulation times.
The process of checking the connectivity of a deposit also consumes much computational
time because it requires
searching for possible isolated clusters \cite{hoshen} after the random step of an adatom.
Moreover, the time for this checking increases with film thickness.

For those reasons, our simulations were restricted to two dimensions and relatively
low values of $R$, ranging from ${10}^1$ to ${10}^4$.
The values of $\epsilon$ were chosen in the range from $0.01$ to $0.15$, which is
sufficient to identify a transition in the porosity scaling.
For $R$ ranging from  $10$ to  ${10}^3$, the lateral size of the deposits
and the number of deposited layers is $400$.
For the largest value $R={10}^4$, the lateral size remains the same but the number of
deposited layers is $100$.
Due to the formation of pores, the average film height may significantly exceed the
number of deposited layers, particularly for large values of $\epsilon$ and $R$.

The porosity $P$ is calculated in the middle of the deposits,
with heights varying from $40$ to $\overline{h}-40$, where $\overline{h}$ is the average
height of the outer surface (highest particles in each substrate column).
$P$ is defined as the ratio between the number of empty lattice sites and the
total number of sites in the scanned region.
The choice of the boundary of this region is suitable to eliminate effects of
the substrate and not to reach the outer surface (note, for instance, that the
roughness of the samples is always below $10$ lattice units).
This procedure parallels that proposed in Ref. \protect\cite{giri} for measuring
porosity is ballistic-like deposits, which consists in selecting only points
below the deepest through of the samples.

The surface roughness of the deposits was also measured.
In a deposit with overhangs, the height of a given column is defined as the
height of the topmost particle at that column. This set of heights defines
the outer surface of the deposit.
For calculating the local roughness,
a square box of lateral size $r$ glides along the film surface and, at each position, the  
root-mean-square (rms) height fluctuation of columns inside the box is calculated.
The average rms fluctuation among all box positions and among different configurations
of the deposit at time $t$ is the local roughness $w\left( r,t\right)$.
The global roughness $W\left( t\right)$ is measured in the full system size $L$, i. e.
$W\left( t\right)=w\left( L,t\right)$.

\section{Structure of the deposits}
\label{porous}

\begin{figure}[!h]
\includegraphics[width=15cm]{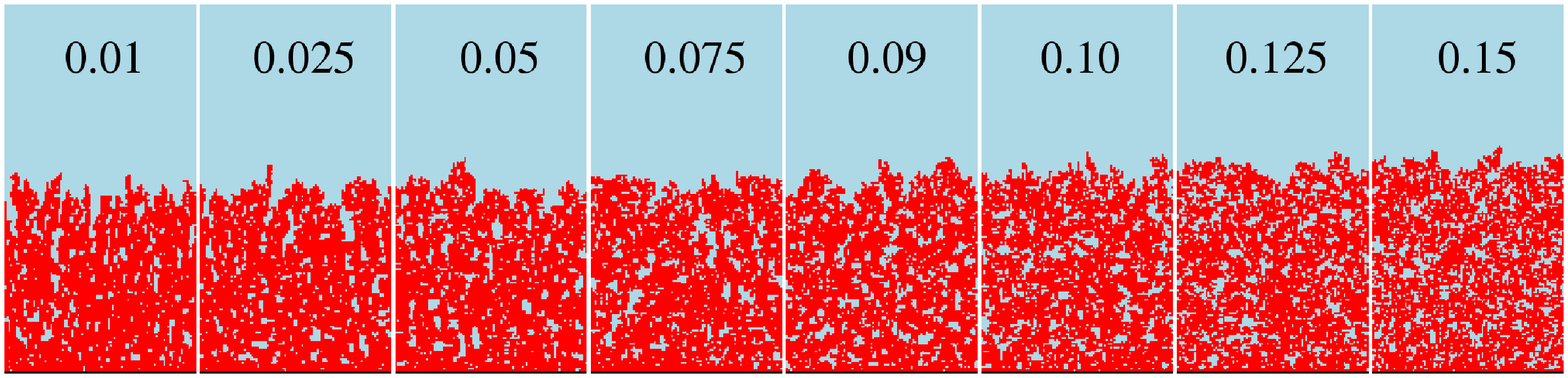}
\caption{(Color online) Snapshots of the layer in $100$-site wide regions,
for $R=10$ and values of \textcolor{red}{$\epsilon$} indicated on the figure.
}
\label{depR10}
\end{figure}

Fig. \ref{depR10} shows deposits grown with $R=10$ and several values of $\epsilon$.
Their porosity slowly increases with $\epsilon$, from $P\approx 0.16$ for
$\epsilon =0.01$ to $P\approx 0.28$ for $\epsilon =0.15$.
This value of $R$ is typical of very low temperature or very large surface energy
($E_s$), leading to a very low adatom mobility in the time scale of deposition of
an atomic layer. Increasing $\epsilon$ does not compensate this overall low mobility.

\begin{figure}[!h]
\includegraphics[width=15cm]{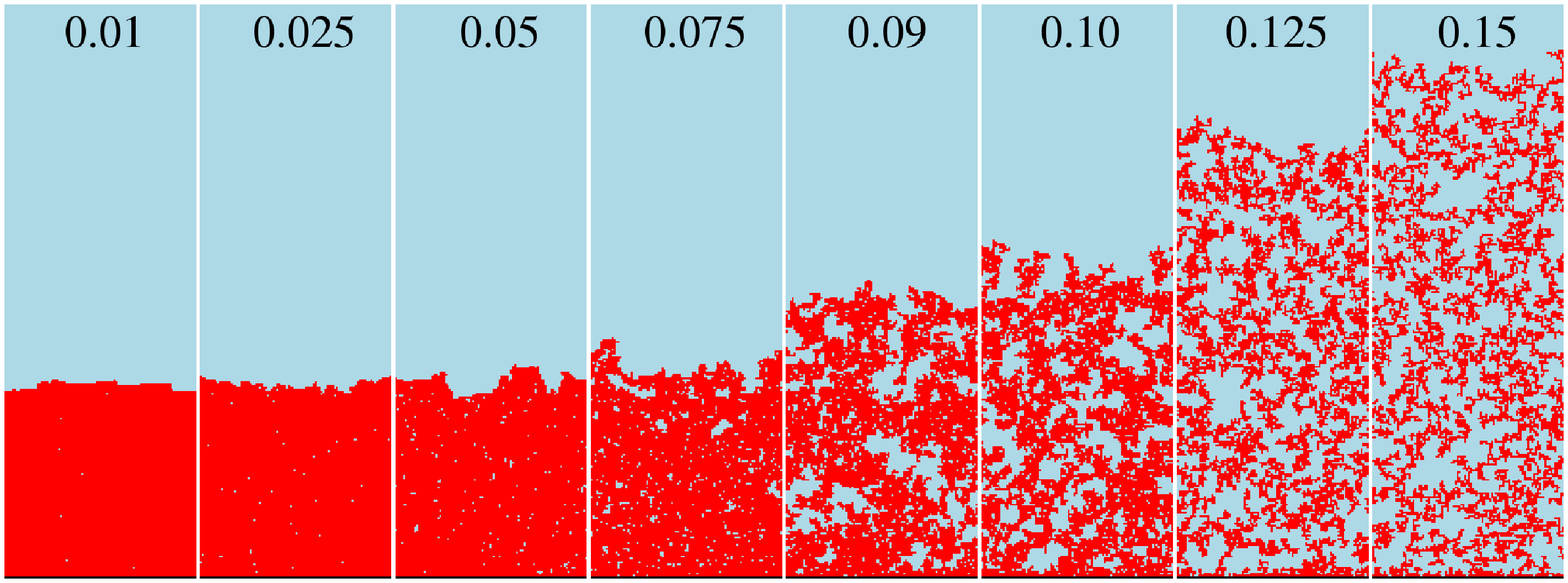}
\caption{(Color online) Snapshots of the layer in $100$-site wide regions,
for $R={10}^4$ and values of \textcolor{red}{$\epsilon$} indicated on the figure.
}
\label{depR10000}
\end{figure}

Fig. \ref{depR10000} shows very different features in deposits grown with $R={10}^4$:
the porosity is negligible for small $\epsilon$ (smaller than $1\%$), but reaches
$P\approx 50\%$ for $\epsilon= 0.15$.

For low $\epsilon$ (low NN binding energy), $R={10}^4$ provides large diffusion
lengths for the adatoms with a small number of NN. Thus, they tend to
aggregate in positions with large numbers of NNs, in which the small $\epsilon$ value
warrants a long residence time. This explains the formation of compact deposits.
However, for $\epsilon\sim 0.1$ or larger, the stability of positions
with many NNs disappears.
All adatoms can easily move, forming overhangs and pores,
with connectivity preserved by constraint on allowed steps.

\begin{figure}[!h]
\includegraphics[width=8cm]{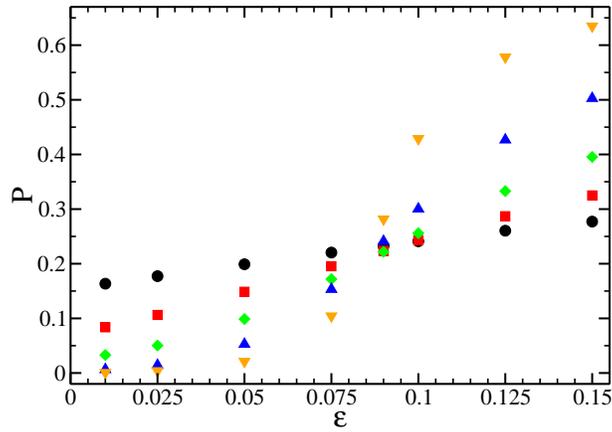}
\caption{(Color online) Porosity as a function of $\epsilon$ for $R=10$ (black circles),
$R=50$ (red squares), $R=200$ (green diamonds), $R={10}^3$ (blue up triangles), and
$R={10}^4$ (yellow down triangles).
}
\label{Pversusepsilon}
\end{figure}

The porosity of deposits for several values of $R$ is shown in Fig. \ref{Pversusepsilon}
as a function of $\epsilon$. It indicates the presence of a dynamic transition in
the large $R$ limit: vanishingly small porosity for $\epsilon\leq 0.075$ and finite
porosity for $\epsilon\geq 0.09$.
The transition point is thus estimated as $0.075<\epsilon_c <0.09$.

This value of $\epsilon_c$ is large, thus the corresponding binding energy $E_b$
is of the same order of thermal energy fluctuations. In terms of that energy,
the transition temperature is estimated as
\begin{equation}
0.386<k_BT_c/E_b<0.415 .
\label{Tc}
\end{equation}
This reminds the large roughening temperatures of thermal equilibrium roughening
(of the same order of melting temperatures) \cite{barabasi,weeks}.

The above discussion considers $R$ and $\epsilon$ as independent parameters.
However, for a given material, $E_s$ and $E_b$ are fixed and related to its
physico-chemical properties. Thus,  $R$ and $\epsilon$ simultaneously change,
being related as
\begin{equation}
R = \frac{h}{F}\epsilon^{E_s/E_b} .
\label{Repsilon}
\end{equation}
This is the basis to investigate the temperature dependence of the porosity.

The data in Fig. \ref{Pversusepsilon} for each value of $R$ was fitted to provide a 
continuous approximation of $P$ as a function of $\epsilon$. The fitting curves
were logistic functions
of the form $f\left( x\right) = a_2/\left[ 1 + \exp{\left( -a_0-a_1 x\right)}\right]$,
with constants $a_0$, $a_1$, and $a_2$ obtained from least squares fits.
For a fixed ratio $E_s/E_b$ and a given value of $R$, $\epsilon$ is calculated by
Eq. (\ref{Repsilon}) and those fits are used to determine $P$.

\begin{figure}[!h]
\includegraphics[width=7cm]{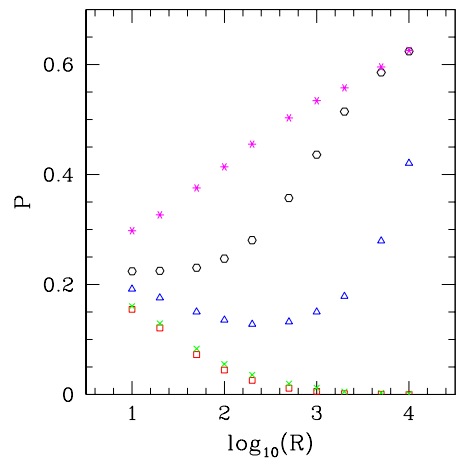}
\caption{(Color online) Porosity as a function of $R$ for several ratios $E_s/E_b$:
$1$ (red squares), $5$ (green crosses), $8$ (blue triangles), $10$
(black hexagons), and $15$ (magenta asterisks).
}
\label{PversusR}
\end{figure}

Fig. \ref{PversusR} shows $P$ versus $R$ for five values of $E_s/E_b$, with fixed
$h/F={10}^{12}$. Some temperature values are also indicated.
For a given ratio $E_s/E_b$, $P$ changes from
a decreasing to an increasing function of temperature at a transition
point, which is the minimum of the correspondig curve
in that plot. For the smallest values of $E_s/E_b$ in Fig. \ref{PversusR},
extrapolation of the results in Eqs. (\ref{Tc}) and (\ref{Repsilon}) indicates that 
the transition occurs only for very large $R$. For instance, for
$E_s=E_b$, the transition value is estimated as $R_c\sim {10}^{11}$;
for $E_s/E_b=5$, the transition value drops to
$R_c\sim {10}^{6}$ (if $E_s=0.5 eV$ and $E_b=0.1 eV$, it gives $T_c\sim 450 K$).

For $R<R_c$ (or $T<T_c$), the decrease of porosity with the temperature
is qualitatively similar to results of related models
\cite{hu2009,zhang2010,sanchez}. 
In Ref. \protect\cite{hu2009}, the activation energy $0.6eV$
and maximal temperature $700K$ give maximal values
of $\epsilon$ of order ${10}^{-4}$, which is in the low temperature regime, even
in a triangular lattice structure. The same range of parameters were considered
in Ref. \protect\cite{zhang2010}. Ref. \protect\cite{sanchez} showed formation
of compact branches in the high temperature growth of some samples, but this was
an effect of a particular lattice structure that suppressed some atomic steps,
leading to an unstable growth. Similar mechanisms are not present in our model.

The nontrivial result of this model is the
porosity increase with the temperature for
$R>R_c$ ($T>T_c$), which was not shown in previous works.
In the transition point, $\epsilon$ is not very small,
thus the hopping rates have a relatively weak dependence on the number of NN.
This leads to a high disorder in the distribution of adatom position, subject to the
constraint of the deposit being connected.
This interpretation antecipates the relation with random percolation,
to be discussed in detail in Secs. \ref{poredistribution} and \ref{roughness}.

\section{Pore size distribution}
\label{poredistribution}

\begin{figure}[!h]
\includegraphics[width=8cm]{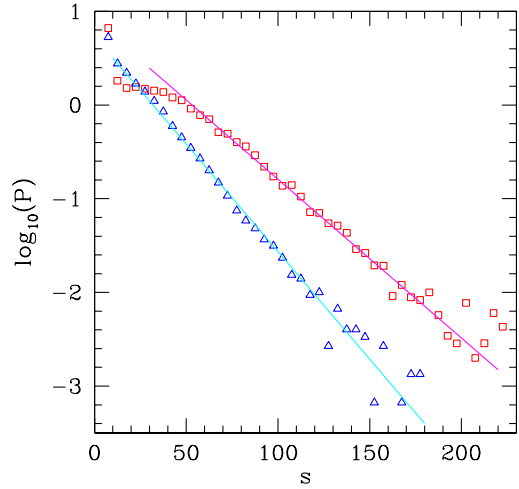}
\caption{(Color online) Pore size distribution below the transition point, for $R={10}^3$
and $\epsilon=0.01$ (triangles) and $\epsilon =0.025$ (squares) (data not normalized).
Solid lines are least squares fits of the data in the approximately linear regions.
}
\label{distbelow}
\end{figure}

Fig. \ref{distbelow} shows the pore size distributions of deposits grown with large $R$
and with $\epsilon$ well below $\epsilon_c$.
For large pore sizes, they have exponential decays.
This is what is typically expected in systems far from a critical point.
Also note that the slopes of the fits in Fig. \ref{distbelow} decrease as $\epsilon$
increases, indicating that the average pore size increased.

\begin{figure}[!h]
\includegraphics[width=8cm]{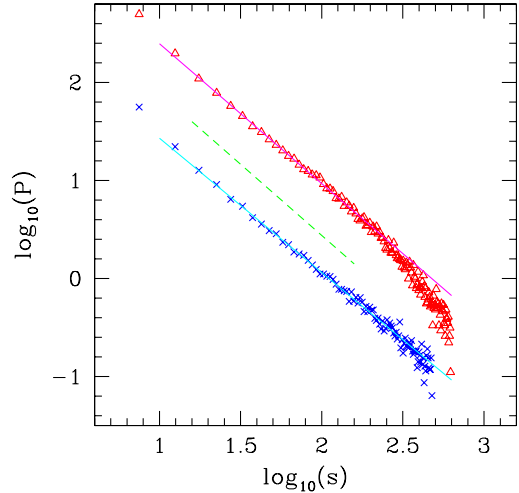}
\caption{(Color online) Pore size distribution (not normalized) above the transition
point, for $\left( R={10}^3,\epsilon=0.125\right)$ (red triangles) and
$\left( R={10}^3\epsilon=0.15\right)$ (blue crosses). The linear fits (solid lines)
have slopes $-1.43$ and $-1.37$, respectively. The dashed line has slope $-1.45$.
}
\label{distabove}
\end{figure}

Fig. \ref{distabove} shows the pore size distribution of deposits grown with large $R$ and
two values of $\epsilon$ well above $\epsilon_c$. A power law decay is observed:
\begin{equation}
P\left( s\right) \sim s^{-\Delta} \qquad ,\qquad 1.35\leq \Delta\leq 1.45 .
\label{dist}
\end{equation}
Here, the error bar was obtained by fitting different ranges of the data for both
values of $\epsilon$ shown in Fig. \ref{distabove}.
This result contrasts to  equilibrium phase transitions \cite{binney}, in which
power law decays are present only at the critical point (or may be observed in a narrow
region around the critical point in finite-size samples).

In the critical point, the rapid adatom dynamics, of diffusive nature, tends to spread atoms
upwards, i. e. to the empty region above the outer surface.
This movement is only constrained by the connectivity condition, so that deposited
atoms always form a cluster connected to the substrate.
This suggests the structure of a critical percolation cluster.

Above the transition point, the adatom dynamics is even faster than that in the
critical point. This favors random adatom distribution and stretches the cluster
in the vertical direction under the connectivity constraint. Thus, we understand that
the cluster is also set into a critical percolation structure by the dynamics.
Due to the external adatom flux, this dynamics is certainly far  from equilibrium.
Consequently, its features may differ from those of equilibrium critical phenomena.
However, as we explain below, the exponents are related to (statistical equilibrium)
percolation exponents.

The critical percolation cluster can be divided in two parts: dead ends, which are isolated
branches, and the backbone, which is the part that carries stress or current if the borders
are mechanically or electrically excited \cite{stauffer}.
Most of the mass of the cluster is in the dead ends.
For this reason, while the fractal dimension of the cluster is $d_F=91/48\approx 1.89$
in two dimensions, the fractal dimension of the backbone is estimated as
$d_B= 1.6432\pm 0.0008$ \cite{grassberger}.

The percolation backbone is frequently modelled as a system of links and blobs
\cite{herrmann,barthelemy}.
The number of blobs of size $b$ in a box with fixed lateral size is
\begin{equation}
n\left( b\right) \sim b^{-\tilde\tau} ,
\label{nb}
\end{equation}
with
\begin{equation}
\tilde\tau = 1 + d_r/d_B ,
\label{tau}
\end{equation}
where $d_r=1/\nu$ is the exponent relating the number of blobs and the system size $L$,
and $\nu=4/3$ is the correlation length exponent of the problem \cite{herrmann,coniglio}.
This gives $\tilde\tau\approx 1.45$.

The internal pores of the deposit are the pores surrounded by the blobs of this connected
cluster. The blob size $b$ is the number of particles (adatoms in our model) in the blob
perimeter. The pore size $s$ of our model is the area (number of empty sites)
enclosed by the blob.
Self-similarity implies that the large blobs have the same shape of the full backbone.
If the blob occupies a region of linear size $r$, then the area and the perimeter are
related as
\begin{equation}
s\sim b\sim r^{d_B} .
\label{sb}
\end{equation}
For this reason, one expects the same decay in the distributions (\ref{dist}) and (\ref{nb}),
which gives
\begin{equation}
\Delta = \tilde\tau .
\label{Deltatheoretical}
\end{equation}

For comparison, Fig. \ref{distabove} illustrates a decay with slope $\tilde\tau$.
Our estimates of $\Delta$ are slightly smaller than this value.
Possibly, this is related to the compact regions inside the largest pores,
which gives an effective dimension of the area $s$ [Eq. (\ref{sb})]
closer to $2$ (consequently larger than $d_B$). 
It can be easily shown that the increase in this effective dimension leads to
effective exponent $\Delta$ smaller than the predicted value.

In Sec. \ref{simulationroughness}, we will show that the roughness scaling gives additional
support to the interpretation that the structure of the deposits above the critical
point is dominated by a percolation backbone.

\section{Surface roughness scaling}
\label{roughness}

\subsection{Basics of kinetic roughening and universality classes}
\label{roughening}

In systems with normal roughening (in opposition to anomalous roughening \cite{ramasco}),
the expected scaling of the local roughness in large substrates is
\begin{equation}
w\left( r,t\right) \approx r^{\alpha} g{\left( \frac{r}{t^{1/z}}\right)} ,
\label{fvlocal}
\end{equation}
where $\alpha$ and $z$ are the roughness and dynamic exponents, respectively, and
$g$ is a scaling function. For $x\equiv r/t^{1/z} \ll 1$ (small box sizes),
$g(x)$ is constant and $w\sim r^\alpha$;
for $x\gg 1$ (large box sizes), the local roughness converges to the global one,
$W\left( t\right)$, which scales as
\begin{equation}
W\left( t\right) \sim t^{\beta} ,
\label{defbeta}
\end{equation}
with $\beta = \alpha /z$ called growth exponent
(see Ref. \protect\cite{chamereis} for a discussion of roughness scaling in
several growth models).

Roughening dominated by adatom surface diffusion is generally believed to be
described by fourth order stochastic growth equations in the continuous limit
\cite{barabasi,krug}:
\begin{equation}
{{\partial h}\over{\partial t}} = \nu_4{\nabla}^4 h +
\lambda_{4} {\nabla}^2 {\left( \nabla h\right) }^2 + \eta (\vec{x},t) ,
\label{vlds}
\end{equation}
where $h(\vec{x},t)$ is the height at position $\vec{x}$ and time $t$ in a $d$-dimensional
substrate, $\nu_4$ and $\lambda_{4}$ are constants and $\eta$ is a Gaussian (nonconservative)
noise [a constant external particle flux is ommited from Eq. (\ref{vlds})].
For $\lambda_4 =0$, Eq. (\ref{vlds}) is linear and is usually called Mullins-Herring (MH)
equation \cite{mh}. The nonlinear form ($\lambda_4 \neq0$) is known as
Villain-Lai-Das Sarma (VLDS) \cite{villain,laidassarma} equation.
In $d=1$, the best estimates of VLDS exponents are obtained from conserved restricted
solid-on-solid models: $\alpha =0.94\pm 0.02$, $z=2.88\pm 0.04$,
and $\beta =0.326\pm 0.011$ \cite{crsosreis}.

In $d=1$ and $d=2$, renormalization studies \cite{haselwandter} showed that the CV model
belongs to the VLDS class. However, simulations show significant scaling corrections
in $d=1$ \cite{tamborenea,lanczycki,kotrla1996,meng} and $d=2$ \cite{wilby}.

On the other hand, most growth models that produce porous deposits are extensions of BD
and are represented by the Kardar-Parisi-Zhang (KPZ) equation  \cite{kpz}
in the hydrodynamic limit:
\begin{equation}
{{\partial h}\over{\partial t}} = \nu_2{\nabla}^2 h +
\lambda_{2} {\left( \nabla h\right) }^2 + \eta (\vec{x},t) ,
\label{kpz}
\end{equation}
with $\nu_2$ and $\lambda_2$ constant.
In $d=1$, KPZ exponents are $\alpha =1/2$, $z=3/2$, and $\beta =1/3$.

The case $\lambda_2=0$ in Eq. (\ref{kpz}) is the Edwards-Wilkinson \cite{ew} equation,
with $\alpha =0.5$ and $\beta=0.25$ in $d=1$.
Refs. \protect\cite{hu2009,zhang2010} proposed that their activated surface diffusion
models were in the EW universality class.

\subsection{Simulation results}
\label{simulationroughness}

\begin{figure}[!h]
\includegraphics[width=8cm]{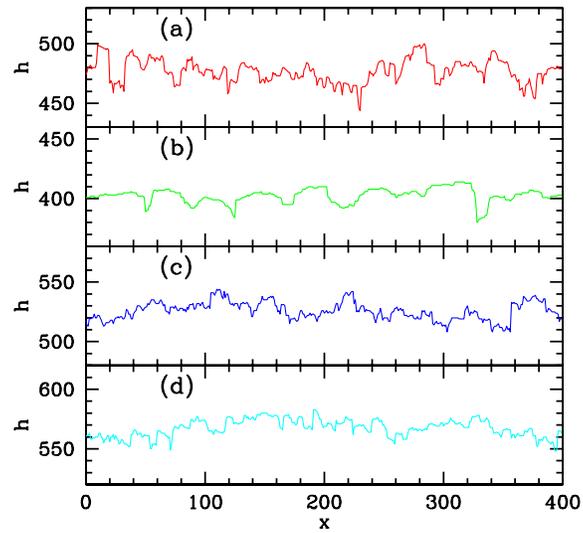}
\caption{(Color online) Outer interface profiles after deposition of $400$ layers
with (a) $R=10$, $\epsilon =0.01$, (b) $R={10}^3$, $\epsilon =0.01$,
(c) $R=10$, $\epsilon =0.1$, (d) $R={10}^3$, $\epsilon =0.1$. 
}
\label{profile}
\end{figure}

Fig. \ref{profile} shows the outer surface profiles of
deposits grown with four pairs of parameters ($R$, $\epsilon$)
after deposition of $400$ layers.

For small $\epsilon$ (Figs. \ref{profile}a and \ref{profile}b; $\epsilon=0.01$),
the interfaces show mounds separated by narrow deep valleys.
These features are more prominent for large
$R$ (Fig. \ref{profile}b), in which the mounds have small roughness and
large widths. This is related to the large diffusion lengths of isolated adatoms in
terraces. In this situation, the porosity is very small.

These trends are similar to the CV model and related collective diffusion models
with solid-on-solid structure.
Some interfaces generated by those models are shown in Refs.
\protect\cite{tamborenea,lanczycki,kotrla1996}:
for low temperatures (small $R$ and very small $\epsilon$), the interfaces are rough,
with rounded peaks separated by sharp deep valleys;
as the temperature increases (large $R$ and $\epsilon$ not very small),
the width of the peaks increase and the roughness decreases; for high temperatures,
the outer surface is very smooth.

Different features are observed above the transition
temperature. Figs. \ref{profile}c and \ref{profile}d, for $\epsilon=0.1$,
show interfaces rough at short and large lengthscales, with narrow mounds
for small and large $R$.

\begin{figure}[!h]
\includegraphics[width=10cm]{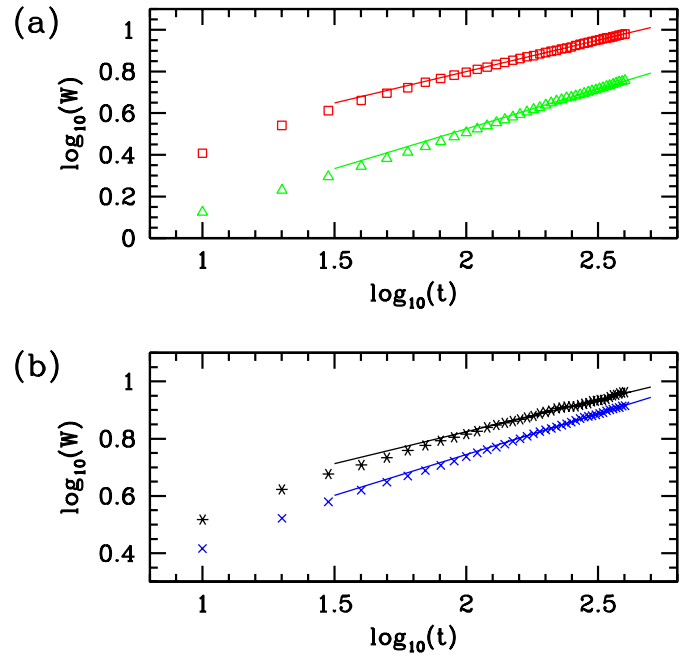}
\caption{(Color online) Global roughness as a function of time and corresponding least squares fits
(solid lines) for:
(a) $\epsilon=0.01$, $R=10$ (red squares; slope $0.30$) and $R={10}^3$ (green triangles; slope $0.38$);
(b) $\epsilon =0.1$, $R=10$ (blue crosses; slope $0.29$) and $R={10}^3$ (black asterisks;
slope $0.22$).}
\label{wglobal}
\end{figure}

In Fig. \ref{wglobal}a, we show the time evolution of the global roughness for deposits
grown with two values of $R$ below the transition point ($\epsilon = 0.01$).
The interface smoothens as $R$ increases, similarly to the CV model.
For $R=10$, the slope of the fit in Fig. \ref{wglobal}a is near the VLDS exponent
$\beta$.
For $R={10}^3$, the slope of the fit is near the exponent $\beta$ of the MH class
\cite{barabasi,mh}. The MH exponents are also observed in some simulations 
of the CV model for large $R$ \cite{tamborenea,lanczycki,kotrla1996} and are
interpreted as an effect of a long crossover to VLDS scaling
\cite{haselwandter}. 

In Fig. \ref{wglobal}b, we show the time evolution of the global roughness for deposits
grown above the transition point ($\epsilon = 0.1$).
In this case, increasing $R$ slightly increases
the roughness, in contrast to the trend of solid-on-solid models.
The slopes of the linear fits are closer to the EW
exponent $\beta=0.25$. However, long range correlations are present in that case
due to the no-desorption condition. Thus, the description of the interface
evolution by local growth equations is expected to fail.

\begin{figure}[!h]
\includegraphics[width=8cm]{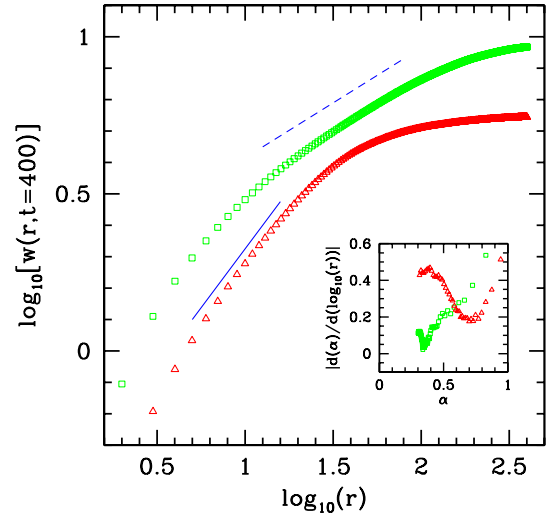}
\caption{(Color online) Local roughness of deposits grown with $R={10}^3$
and $\epsilon=0.01$ (red triangles) and $\epsilon =0.1$ (green squares).
Solid and dashed lines are drawn to guide the eye and
have slopes $0.75$ and $0.35$, respectively.
Inset: $|d\alpha/d\log{r}|$ as a function of $\alpha$, with symbols corresponding
to the same parameters of the main plot.
}
\label{wlocal}
\end{figure}

In Fig. \ref{wlocal}, the local roughness $w\left( r,t=400\right)$ is plotted
as a function of the box size $r$ for deposits grown above and below the transition point
with $R={10}^3$.

The dominant slopes of those plots are calculated by a
method proposed in Ref. \protect\cite{chamereis}.
First, the effective exponent $\alpha{\left( r,t\right)}$ is defined as the local
slope of the $\log{w\left( r,t\right)}\times\log{r}$ plot.
Then, the dominant slope is defined as the exponent $\alpha$ that minimizes
$\left|\frac{d\alpha{\left( r,t\right)}}{d\log{r}}\right|$.
The inset of Fig. \ref{wlocal} shows  $|d\alpha/d\log{r}|$ as a function of
$\alpha$ calculated using this procedure, showing 
minima at $\alpha\approx 0.75$ for $\epsilon =0.01$ and 
$\alpha\approx 0.35$ for $\epsilon =0.1$.
In the main plot of Fig. \ref{wlocal}, dashed lines have these slopes.

The initial slope $\approx 0.75$ is dominant for $\epsilon =0.01$.
This is interpreted as an effect of the grainy interface structure
shown in Figs. \ref{profile}a,b,
with wide and rough plateaus separated by narrow and deep valleys.
For square surface blocks, the expected initial slope is $1$ \cite{graos}.
For rounded surface blocks, that slope is slightly smaller, typically between
$0.7$ and $1$, which is the present case \cite{grainshape}.
This is a geometric effect, with no relation to kinetic roughening.

The roughness saturates immediately after that initial regime,
thus we cannot obtain a reliable estimate of the roughness exponent
for $\epsilon =0.01$.
The VLDS roughness exponent is also large ($\alpha\approx 0.94$
\cite{crsosreis}), thus a possible crossover may be hidden by the
large initial slope.
A long time crossover to KPZ scaling is also plausible because
a small KPZ nonlinearity may be generated by the formation of overhangs and pores
\cite{pellegrini,kikkinides,bbd,perez,yu,bbdflavio,khanin,banerjee}.
Also recall that slow crossovers to KPZ scaling are observed in the films with grainy
interfaces of Refs. \protect\cite{graos,grainshape}, which resemble 
those in Figs. \ref{profile}a,c,d.

For $\epsilon =0.1$, a scaling region appears in Fig. \ref{wlocal}
with slope $\approx 0.35$, which
differs from the roughness exponents $\alpha$ of EW, KPZ, MH, and VLDS classes.
This is expected due to the long-range correlation introduced by the
no-desorption conditions. On the other hand, 
the fractal dimension corresponding to this roughness (Hurst) exponent
is $d_f=2-\alpha=1.65$, which is in very good agreement with
the dimension of the percolation backbone in two dimensions,
$d_B\approx 1.6432\pm 0.0008$ \cite{grassberger}.

This result shows that the upper parts of the deposits above the transition point
are also dominated by the structure (link-node-blob) of a percolation backbone.
The contribution of the dead ends of the full percolation cluster is consequently
negligible, which is reasonable for the scaling of the roughness at long wavelengths
(large box size $r$).
As explained in Ref. \protect\cite{barabasi}, the above self-similarity interpretation
is also restricted to $r\ll L$, which is the case of the range of $r$ fitting the slope
$0.35$ in Fig. \ref{wlocal}.

We tried to improve the analysis of surface roughness scaling with alternative
approaches, but did not succeed. First, we used a method recently proposed
to reduce the corrections to scaling in ballistic-like deposition models 
\cite{baltiago}, but found no significant change in exponents $\beta$ and $\alpha$.
Secondly, we measured roughness distributions in the growth regime.
However, they had many fluctuations due to the small number of configurations
calculated in a lattice size not very large, hindering a reliable comparison
with distributions of KPZ, VLDS, and other growth classes.

\section{Conclusion}
\label{conclusion}

We studied a film growth model in two dimensions with temperature activated diffusion
of adsorbed particles. The diffusion coefficient has the same form of
the Clarke-Vvedensky model, but formation of overhangs and pores is allowed
after an adatom step, with a no-desorption condition that rejects steps
that lead to disconnection of one or more adatoms from the main deposit.
The model was studied in $1+1$ dimensions due to computational limitations.

Taking the diffusion-to-deposition ratio $R$ and the NN binding parameter $\epsilon$
as independent quantities, we show that a critical value $\epsilon_c$ separates
regimes of low and high porosity. As $R$ increases, the transition between these
regimes becomes steeper as $\epsilon_c$ is crossed, with 
vanishingly small porosity for $\epsilon <\epsilon_c$ and finite porosity for
$\epsilon >\epsilon_c$.

Taking fixed values of activation energies $E_s$ and $E_b$, two regimes are
separated by a transition temperature $T_c$, with porosity decreasing (increasing)
with temperature for $T<T_c$ ($T>T_c$). If $E_s\sim E_b$, the value of $R_c$ is
very large and the transition temperature is of the order of $E_b/k_B$; this
resembles the roughening transition in equilibrium models \cite{barabasi,weeks}.

The deposits grown with large $R$ and $\epsilon >\epsilon_c$ are tuned
into a critical percolation regime. The pore size distribution has a power-law decay,
whose exponent is related to critical percolation exponents.

We also studied the kinetic roughening of the outer interface of the deposits.
Below the transition point, growth exponents near the VLDS values are obtained,
but a crossover to long time KPZ scaling cannot be discarded due to the small excess
velocity provided by the formation of some small pores.
Above the transition point, we obtain a roughness exponent $\alpha\approx 0.35$,
which is related to the fractal structure of the percolation backbone.

Most of the results presented here are also expected if
three-dimensional deposits are grown with this model.
First, the transition in  porosity is related only to the change in the mobility
of adatoms with two or more neighbors. Second, the structure of the deposits
at and above the transition point is also expected to be of a percolation
backbone because this feature is related to the connectivity constraint
and to the large adatom mobility. Third, the kinetic roughening
of the outer surface will follow the same trend: below the transition point,
VLDS scaling is also expected
for the CV model in $2+1$ dimensions \cite{haselwandter,wilby}, with a possible
crossover to KPZ due to pore formation; above the critical
point, the roughness exponent is also expected to be related to the percolation
cluster structure. 
On the other hand, the pore size distribution may change drastically in $2+1$
dimensions because the density of the deposits will be much smaller and the
pore space will probably be infinitely connected (this is the case, for instance,
of ballistic deposits \cite{yu,bbdflavio,khanin}).

Our results suggest that adsorbed molecule diffusion may also be considered to
explain increase of porosity in thin films, in contrast to the usual expectation
that more compact and smoother films are obtained as the temperature increases.
This unusual trend was already observed in electrostatic spray deposition of oxides
\cite{chen,marinha}, but other mechanisms were considered to explain those
features, such as the evaporation of the spray droplets.

\begin{acknowledgments}
This work was partially supported by CNPq and FAPERJ (Brazilian agencies).
\end{acknowledgments}



\vskip 3cm

\end{document}